\documentclass[11pt,a4paper,UKenglish,texlive=2016]{article}
\pdfoutput=1
\usepackage{journal/jinstpub}

\usepackage[biblatex=false]{atlaspackage}

\usepackage{epsfig}
\usepackage{float}
\usepackage{longtable}

\graphicspath{{logos/}{figures/}}

\usepackage{multirow}
\DeclareSIUnit{\sample}{S}
\title{Destructive breakdown studies of irradiated LGADs at beam tests for the ATLAS HGTD}

\makeatletter
\newcommand\footnoteref[1]{\protected@xdef\@thefnmark{\ref{#1}}\@footnotemark}
\makeatother

\author[a]{L. A.~Beresford,}
\author[b]{D. E.~Boumediene,\footnote{\label{note1}Corresponding author.}}
\author[c]{L.~Castillo Garc\'{i}a,\footnoteref{note1}}
\author[b]{L. D.~Corpe,}
\author[d]{M. J.~Da Cunha Sargedas de Sousa,}
\author[e]{H.~El Jarrari,}
\author[f]{A.~Eshkevarvakili,}
\author[c]{C.~Grieco\footnote{now at UCSB},}
\author[c,n]{S.~Grinstein,}
\author[f]{S.~Guindon,}
\author[g]{A.~Howard,\footnoteref{note1}}
\author[g]{G.~Kramberger,}
\author[h]{O.~Kurdysh,}
\author[i]{R.~Mazini,}
\author[j]{M.~Missio,}
\author[k]{M.~Morenas,}
\author[b]{O.~Perrin,}
\author[l]{V.~Raskina,}
\author[m]{G.~Saito,}
\author[l]{S.~Trincaz-Duvoid,}

\affiliation[a]{Deutsches Elektronen-Synchrotron (DESY), Notkestraße 85, 22607 Hamburg, Germany}
\affiliation[b]{Laboratoire de Physique de Clermont-Ferrand (LPC), Universite
Clermont Auvergne, CNRS/IN2P3, Campus Universitaire des Cézeaux, 4 Avenue Blaise Pascal, 63178 Aubière Cedex, France}
\affiliation[c]{Institut de F\'{i}sica d'Altes Energies (IFAE), The Barcelona Institute of Science and Technology (BIST), \\Carrer Can Magrans s/n, Edifici Cn, Campus UAB, E-08193 Bellaterra (Barcelona), Spain}
\affiliation[d]{Department of Modern Physics and State Key Laboratory of Particle Detection and Electronics, University of Science and Technology of China (USTC), 96 JinZhai Road Baohe District, Hefei, Anhui, 230026, China}
\affiliation[e]{Universit\'{e} Mohammed V de Rabat, Avenue des Nations Unies, Agdal, Rabat, Morocco}
\affiliation[f]{Conseil Europ\'{e}en pour la Recherche Nucl\'{e}aire (CERN), Esplanade des Particules 1, CH-1211 Meyrin, Switzerland}
\affiliation[g]{Jo\v{z}ef Stefan Institute (JSI), Jamova cesta 39, 1000 Ljubljana, Slovenia}
\affiliation[h]{Laboratoire de Physique des 2 Infinis Irène Joliot Curie (IJCLab), 15 Rue Georges Clemenceau, 91400 Orsay, France}
\affiliation[i]{Academia Sinica, 128, Section 2, Academia Road, Nangang District, Taipei City, Taiwan 115}
\affiliation[j]{Institute for Mathematics, Astrophysics and Particle Physics, Radboud University/Nikhef, Nijmegen, P.O. Box 9010, 6500 GL Nijmegen, Netherlands}
\affiliation[k]{Omega, Ecole Polytechnique, Palaiseau, France}
\affiliation[l]{Laboratoire de Physique Nucl\'{e}aire et de Hautes Energies (LPNHE), Sorbonne Universit\'{e}, Universit\'{e} de Paris, CNRS/IN2P3, Paris, France}
\affiliation[m]{Instituto de Fisica, Universidade de Sao Paulo, Sao Paulo, Brazil}
\affiliation[n]{Instituci\'{o} Catalana de Recerca i Estudis Avan{\c c}ats (ICREA), Passeig de Llu\'{i}s Companys, 23, 08010 Barcelona, Spain}

\emailAdd{djamel.boumediene@cern.ch, lucia.castillo.garcia@cern.ch, alissa.howard@ijs.si}

\abstract{%

In the past years, it has been observed at several beam test campaigns that
irradiated LGAD sensors break with a typical star shaped burn mark when
operated at voltages much lower than those at which they were safely operated
during laboratory tests. The study presented in this paper was designed to
determine the safe operating voltage that these sensors can withstand. Many
irradiated sensors from various producers were tested in two test beam
facilities, DESY (Hamburg) and CERN-SPS (Geneva), as part of ATLAS High
Granularity Timing Detector (HGTD) beam tests. The samples were placed in the
beam and kept under bias over a long period of time in order to reach a high
number of particles crossing each sensor. Both beam tests lead to a similar
conclusion, that these destructive events begin to occur when the average
electric field in the sensor becomes larger than
\SI{12}{\volt\per\micro\metre}. } 


\arxivnumber{}

\begin{document}

\maketitle

\section{Introduction}
\label{sec:intro}
During the high-luminosity phase of the LHC (HL-LHC), the primary interactions
created in proton-proton collisions are accompanied by a large number of
zero and minimum bias interactions (pileup). 
In order to mitigate the adverse effects of pileup, the ATLAS experiment will
install the High Granularity Timing Detector (HGTD)~\cite{ATLAS-TDR-31}. In
particular, the high-precision track timing information from HGTD 
will allow for the proton-proton collisions to be distinguished in time. HGTD
will use Low Gain Avalanche Detectors 
(LGADs)~\cite{PELLEGRINI201412,MarTorino}. LGADs provide low gain, typically
to the order of several tens, by using a n$^{++}$-p$^{+}$-p-p$^{++}$ structure
to create electric fields high enough for impact ionization to occur at the
n$^{++}$-p$^{+}$ junction. The doping concentration and profile shape of the
gain layer impacts the gain factor of the detector.

The harsh radiation environment at the HL-LHC, particularly in the forward region (corresponding to a pseudorapidity\footnote{
  The ATLAS experiment uses a right-handed coordinate system with its origin at the nominal interaction point (IP) in the center of the detector, and the $z$-axis along the beam line. The $x$-axis points from the IP to the center of the LHC ring, and the
 $y$-axis points upwards. Cylindrical coordinates $(r, \phi)$ are used in the
 transverse plane, $\phi$ being the azimuthal angle around the
 $z$-axis. Observables labelled ``transverse'' are projected onto the
 $x$~-~$y$ plane. The pseudorapidity is defined in terms of the polar angle
 $\theta$ as $\eta=-\ln\tan \theta/2$.}, $\eta$, range of 2.8<$\eta$<4.0)
where HGTD will operate, will lead to the deterioration of LGAD
performances. The main concern is the so-called initial acceptor
removal~\cite{LGADIrrad}, which reduces the gain and requires an increase of
bias voltage to compensate for the loss of acceptors in the gain layer. By the
end of their lifetime at HGTD the most exposed sensors will have received a
\SI{1}{\mega\electronvolt} neutron equivalent fluence of around
$\Phi_{eq}=2.5\times10^{15}$~n$_{eq}$/cm$^{2}$. This would require operation of
sensors at voltages exceeding \SI{700}{\volt} at \SI{-30}{\celsius} for
a sensor active area thickness of \SI{50}{\micro\metre}. It has been shown in
the laboratory using strontium-90 electrons that in such conditions enough
charge is collected and the targeted time resolution of 35 (70)\,ps at the start (end) of their lifetime.
The minimal bias voltage under which each sensor reaches the 4 fC charge threshold was first measured in the laboratory using a strontium-90 source.

\begin{figure}[t]
\centering
\begin{tabular}{cc}
\epsfig{file=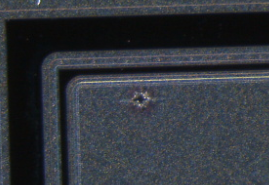,width=0.436\linewidth,clip=} & \epsfig{file=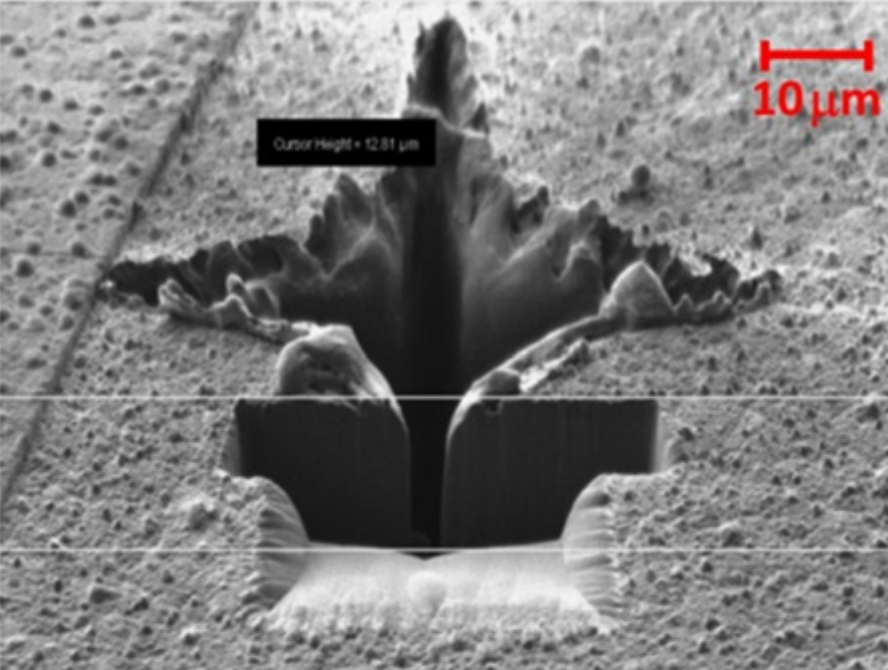,width=0.4\linewidth,clip=} \\
(a) & (b) 
\end{tabular}
\caption{(a) Typical star shaped burn mark observed in beam test sensor
  mortalities due to SEB. Sensor pictured here is from run HPK-P2 wafer 31 irradiated to 2.5$\times10^{15}$~n$_{eq}$/cm$^{2}$, whose mortality occurred at the DESY beam test. (b) Microscopic photograph of a typical burn mark, observed in ATLAS proton beam tests at Fermilab in 2018 in a CNM LDA35 sensor (courtesy of CNM)~\cite{CNMphoto}.}
\label{fi:StarMark}
\end{figure}

 The HGTD beam test campaigns at CERN-SPS~\cite{spsfacility} (\SI{120}{\giga\electronvolt} pions) and DESY~\cite{desyfacility}
(\SI{3}{\giga\electronvolt} electrons) aimed to reproduce the strontium-90 measurements. Many of the sensors underwent destructive breakdown at voltages that were $\sim\SI{100}{\volt}$ lower than those at which the sensors were successfully operated in laboratory tests. 
A typical star shape burn mark, see figure~\ref{fi:StarMark}, appeared in the location of the particle hitting the sensors.  

\begin{figure}[t]
\centering
\includegraphics[width=0.75\textwidth]{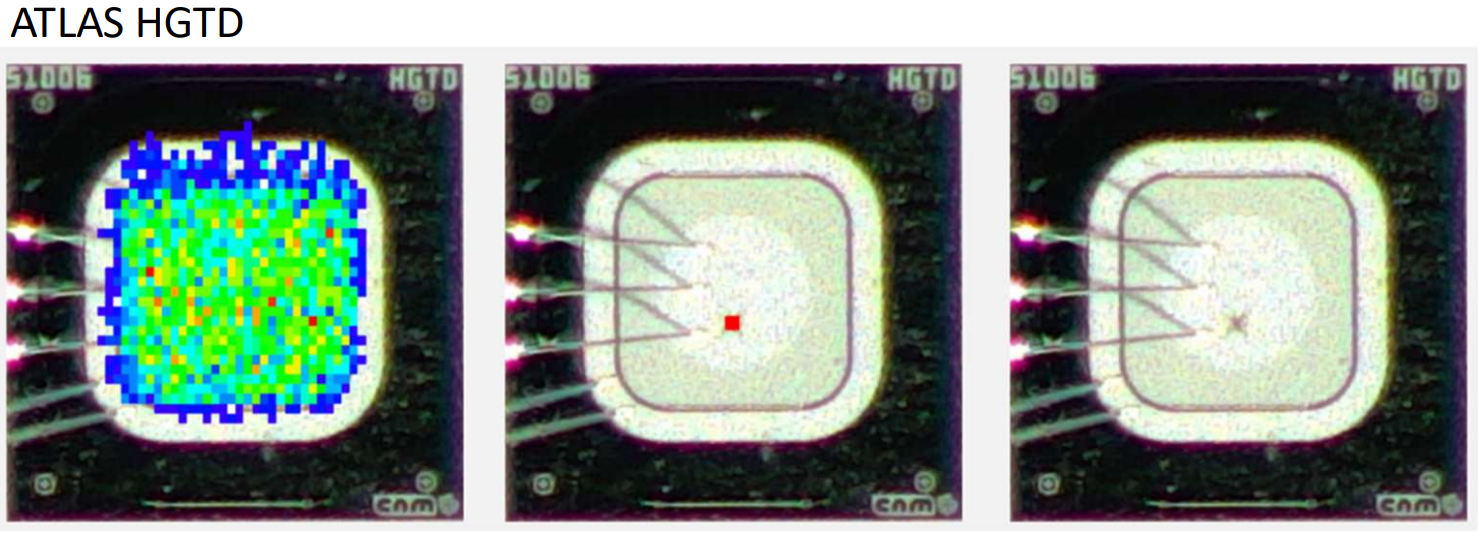}
\caption{The right plot shows a typical single event burnout mark from a 2019 DESY beam test with \SI{5}{\giga\electronvolt} electrons. The reconstructed track in the destructive event pointed to the location of the burn mark (middle and right plot). The distribution of all the reconstructed tracks across the detector before destruction is shown in the left plot.}
\label{fi:TrackReconstruct}
\end{figure}

An intensive investigation done within the ATLAS, CMS and RD50 Collaborations~\cite{Kramberger,Heller}
led to the conclusion that a single beam particle hitting the detector is
responsible for the sensor destruction. Both ATLAS and CMS recorded the signal from such events and associated the location of the crater with the hit position reconstructed by the beam telescope. An example is shown in figure~\ref{fi:TrackReconstruct}. Studies with high-power \SI{50}{\femto\second} laser pulses were performed~\cite{SEE} and also confirmed the destructive events at high bias voltages.

The mechanism of the destruction is called single event burnout
(SEB)~\cite{SEB} and will be schematically described in the next
section. However, it was not clear what the parameters determining its
occurrence were. In order to answer that question a large set of sensors from
various producers with different thicknesses of the active material, D, irradiation
levels, annealing stages and gain layer designs, were tested in two different
beam test campaigns at DESY and SPS, in 2021. This paper describes the result of these
studies.

\section{Single event burnout}
\label{sec:seb}
\begin{figure}[t]
\centering
\includegraphics[width=0.80\textwidth]{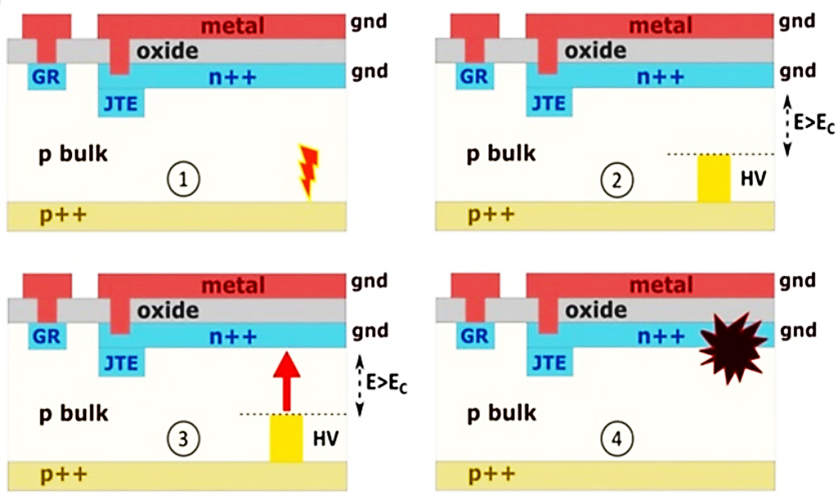}
\caption{Scheme of the SEB mechanism in an LGAD~\cite{SEB}. (1) A large amount of energy is deposited in the sensor. (2) Large carrier density leads to collapse of the field. (3) The HV is brought closer to the pad leading to very high field strength. (4) Avalanche breakdown leads to destruction of the sensor.}
\label{fi:SEBscheme}
\end{figure}

The most probable cause of the sensor mortalities observed is the mechanism
known as SEB. Laboratory tests use a beta beam from a strontium-90 source
which have significantly less energy than the test beams. The maximum energy
of a strontium-90 electron is \SI{2.3}{\mega\electronvolt} which sets the
maximum deposited charge in the LGAD sensor. 
On the other hand, the deposited charge from a high-momentum particle beam
in the active zone of the LGAD detector
can be much larger. 
According to GEANT4~\cite{geant4} simulations up to
\SI{100}{\mega\electronvolt} can be deposited by a single charged
hadron~\cite{Heller,Huhtinen:687409}. This leads to the generation of a large density of
carriers. The screening effect created by such a large carrier density then
prevents carriers from being swept away, leading to a change in 
the local resistivity such that the sensor becomes conductive~\cite{SEB}. The
field collapses in the region of high free carrier 
density, leading to an increased voltage drop in the region where density is lower. The increase of the field there leads to avalanche breakdown (field exceeds the critical field). The charge stored on the sensor electrodes as well as on the high voltage (HV) filtering capacitor (typically \SI{10}{\nano\farad}) is discharged through the sensor. The energy available in such a discharge is enough to melt the silicon, creating a crater and damaging the sensor. The breakdown is eventually quenched, but the sensor is permanently damaged.
This mechanism is illustrated in figure~\ref{fi:SEBscheme}.

\section{Test beam set-up}
\label{sec:Setup}
\begin{figure}[t]
\centering
\begin{tabular}{cc}
\epsfig{file=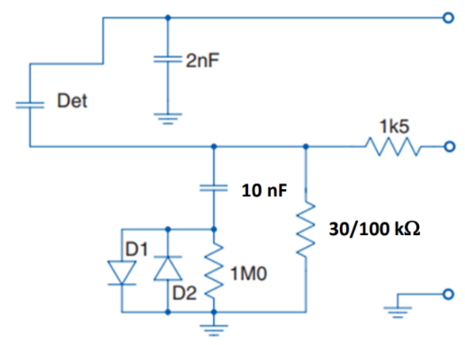,width=0.4\linewidth,clip=} & \epsfig{file=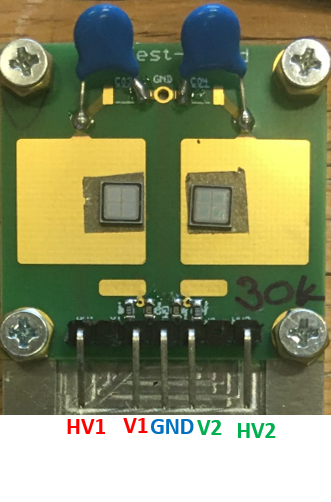,width=0.2\linewidth,clip=} \\
(a) & (b) 
\end{tabular}
\caption{(a) Circuit diagram of a single channel of the PCB. (b) Photograph of the PCB, showing the five connections for: bias applied to the left sensor (HV1), the voltage output of the left sensor (V1), ground (GND), the voltage output of the right sensor (V2), and the bias applied to the right sensor (HV2).}
\label{fi:TB-PCB}
\end{figure}

\begin{figure}[t]
\centering
\includegraphics[width=0.50\textwidth]{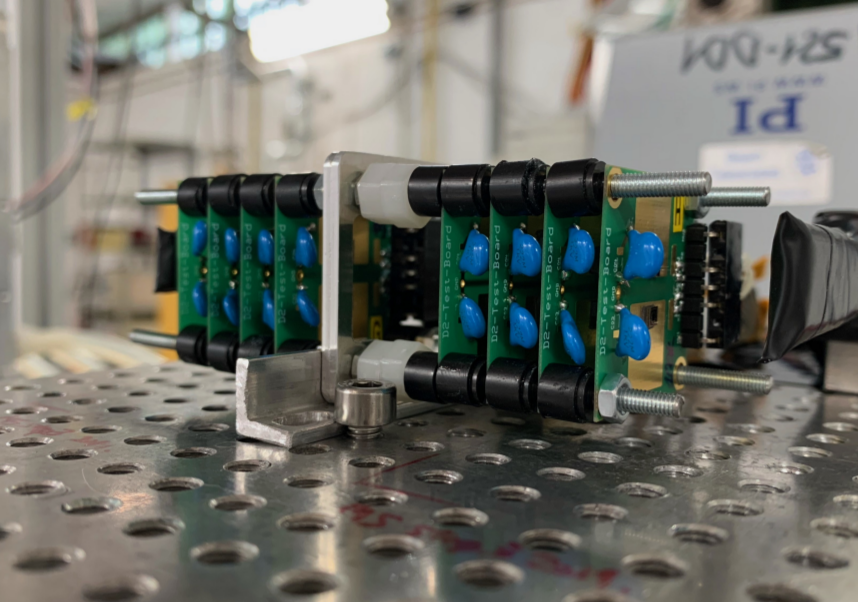}
\caption{The eight PCBs are placed one in front of the other on a frame, and
  placed directly in the beam. A scintillator and PMT placed in front of the
  train of boards record the beam rate. Another scintillator and PMT are
  placed behind the train. The coincidence of both was used to align the
  sensors in the beam.} 
\label{fi:EoL_Train}
\end{figure}

The SEB events were initially observed in sensors being operated in test beam
lines where occasionally the signals from the fatal events were recorded.
In order to maximise the number of sensors that can be tested at a
given beam test campaign, a printed circuit board (PCB) was designed
that could host two sensors of different sizes. Figure~\ref{fi:TB-PCB} shows
the circuit diagram of the boards, and the five connections to the boards. The
current was measured as a voltage drop on the bias resistor. The resistor is
either \SI{30}{\kilo\ohm} or \SI{100}{\kilo\ohm} based on the size of the
sensors; smaller resistance for larger-sized sensors. A \SI{2}{\nano\farad}
capacitor was used to simulate the full sensor capacitance, while the
circuitry in figure~\ref{fi:TB-PCB}a prevented the voltage drop from exceeding
the maximum tolerable at the analog-to-digital converter (ADC) input. Up to eight PCBs, i.e. 16 sensors, were aligned to one another with the use of mechanical rails, as shown in figure~\ref{fi:EoL_Train}. 

A CAEN N472 HV power supply,
 which has four HV channels, was used to apply the bias to the sensors. To
 allow for more than one sensor to be biased by one HV channel the HV was
 first taken to an external box where it is split to allow up to seven
 connections per HV channel. Flat cables carried the HV to each of the PCBs. A
 major benefit of this design is that when a sensor broke it simply needed to
 be disconnected from this box and all the other sensors connected to the same
 HV channel could continue to be operated.

A NI-USB 6001 DAQ system was used to measure the voltage drop on the sensors,
as well as the HV channels, temperature and humidity. This setup was used at
both beam tests.
At DESY dry ice was used to cool down the sensors. Since the temperature
increases as the dry ice evaporates, detailed temperature measurement was kept
throughout testing, which required a dedicated HV channel. 3 HV channels were
used for sensor testing.  The HV was applied only while the temperature was between
\SI{-40}{\degree C} to \SI{-25}{\degree C}. At the SPS the cooling box used a
chiller to keep the temperature stable at \SI{-30}{\degree C}, which is the
planned operating temperature of the HGTD in the ATLAS detector, and so the choice was made to measure 4 HV channels and just monitor the temperature by using an external sensor placed inside the cooling box. 

In both beam tests, a 1~cm$^{2}$ scintillator read out by a photo multiplier tube
(PMT) placed in front of the train of sensors recorded the beam rate, which
allowed for the number of particles crossing through each sensor to be
calculated. This scintillator was used together with a second
scintillator placed behind the train of sensors to align the sensors in the
beam by moving the platform to observe where the coincidence rate was
maximum.

\section{Sensors}
\subsection{LGAD samples}
The ATLAS sensors will consist of 15$\times$15 LGAD arrays of
1.3$\times$1.3~mm$^2$ pads. For the tests performed here other pad
structures are used such as single-pads and 2$\times$2 and 5$\times$5
arrays. A complete list of the sensors sent to the beam tests and their
properties is given in appendix~\ref{A:SensorList}. A set of sensors from
different runs and wafers, produced by HPK\footnote{Hamamatsu Photonics,
  Japan.}, IHEP-IME\footnote{Institute of High Energy Physics and Institute of
  Microelectronics, Chinese Academy of Sciences, China.}, CNM\footnote{Centro
  Nacional de Microelectr\'onica, Spain.}, FBK\footnote{Fondazione Bruno
  Kessler, Italy.}, NDL\footnote{Novel Device Laboratory, China.} and
USTC-IME\footnote{University of Science and Technology of China and Institute
  of Microelectronics, China.}, were chosen to test a wide variety of LGAD
properties. The sensors also vary in size, active material thickness and
in the radiation level they were subjected to, all given in the tables~\ref{tab:sensors1}
and~\ref{tab:sensors2}. Irradiation fluence is given as the
\SI{1}{\mega\electronvolt} neutron equivalent~\cite{RD48}. All of the sensors
were irradiated with reactor neutrons at the Jozef Stefan Institute's TRIGA II
research reactor~\cite{Reactor}. Most of the sensors were then annealed for
\SI{80}{\minute} at \SI{60}{\degree C}, unless otherwise stated. 
Of the sensors tested, USTC-IME-1.1 wafer 11 sensors, all FBK sensors and all
IHEP-IMEv2 sensors are carbon-enriched~\cite{LGADCarbon} while the rest are not.

A few sensors were studied for more specific reasons. The HPK-P1 sensors of
type-3.1 and 3.2 were annealed for an extremely long time, and for this reason
were chosen in order to test if annealing time would have any impact on the
occurrence of mortality. The performance of these type-3.1 and 3.2 sensors
were tested at HGTD beam tests in 2018 and 2019~\cite{HPK3-2018/19}. HPK-P1
sensors of type-1.1 and 1.2 were chosen to give a clearer picture if mortality
was impacted by the sensor active material thickness, as they are thinner than
the more recently produced LGADs at \SI{35}{\micro\metre} versus the more
typical $\sim$\SI{50}{\micro\metre}. Lastly, irradiated Positive Intrinsic
Negative (PIN) diodes were chosen in order to determine if gain had an impact on the mortality.

\subsection{Testing process}
In total, 32 samples were sent to DESY and were measured in two batches of 16
sensors across two weeks. Those that survived the DESY test beam were sent to
the SPS test beam. In total, 42 sensors were sent to SPS and measured in four
batches altogether, again across two weeks. In each case, the sensors were
grouped up in such a way that sensors requiring similar biases to achieve the
\SI{4}{\femto\coulomb} requirement for the HGTD~\cite{ATLAS-TDR-31} were
connected on the same HV channel. The bias of each HV channel was then
increased until it was high enough to reach \SI{4}{\femto\coulomb} for the
best performing sensor. The sensors were left for a significant amount of time
in the beam, so as to acquire a high number of particles crossing the
sensor. At DESY, where the particle rate was approximately
\SI{2}{\kilo\hertz\per\centi\metre\squared} through the entire beam test
period, the sensors were left in the beam for 6 to 8 hours so as to reach a
total of about a million particles crossing each sensor pad in this time. At
the SPS, the rate was less stable but overall significantly higher, reaching up
to more than \SI{6}{\kilo\hertz\per\centi\metre\squared} at times, so most
sensors received many millions of particles per pad. Once at least $10^6$
events had been reached, the bias was increased by about \SI{20}{\volt}. This
was repeated until either the sensor broke or a bias well above 
the one required to achieve the \SI{4}{\femto\coulomb} charge requirement was
reached. Sensor breakdown was identified through a current divergence.

\section{Results and discussion}
\label{sec:Results}
\begin{figure}[t]
\centering
\begin{tabular}{cc}
\epsfig{file=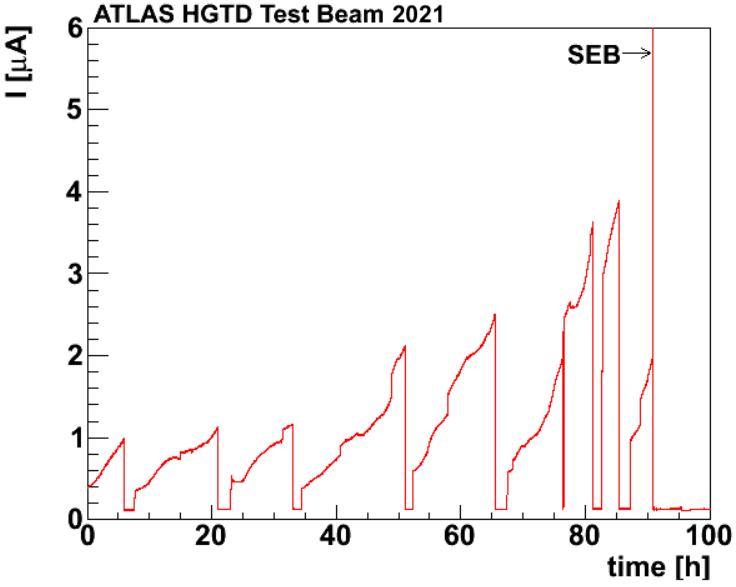,width=0.45\linewidth,clip=} & \epsfig{file=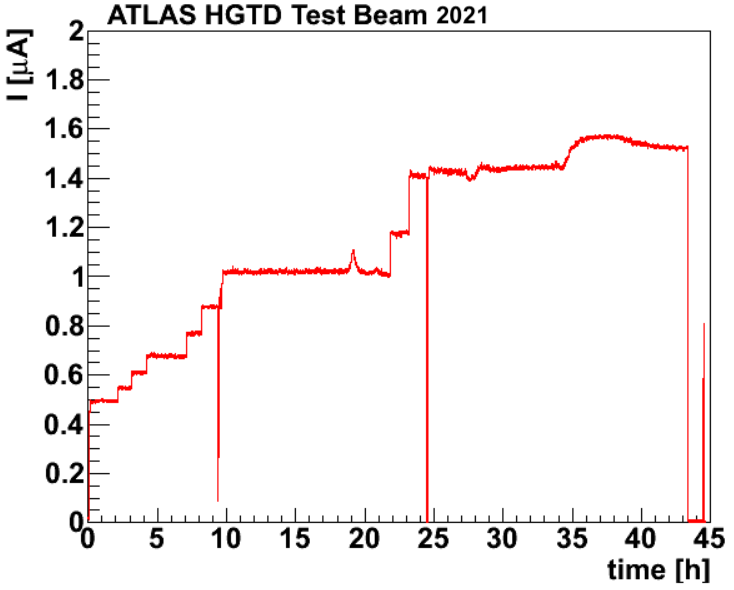,width=0.45\linewidth,clip=} \\
(a) & (b) 
\end{tabular}
\caption{Example of the plots produced from the data obtained at the beam tests. (a) The current of a HPK-P2 W25 SE3-IP4 2$\times$2 sensor irradiated to $1.5\times10^{15}$~n$_{eq}$/cm$^{2}$ measured at DESY. It broke at 90 hours. (b) The current of a IHEP-IMEv2 W4-II 1$\times$3 sensor irradiated to $1.5\times10^{15}$~n$_{eq}$/cm$^{2}$ measured at SPS.}
\label{fi:Current}
\end{figure}

 An example of the plots produced from the recorded data is shown in figure~\ref{fi:Current}. Figure~\ref{fi:Current}a shows the current of a HPK-P2 W25 2$\times$2 sensor as a function of time measured during the DESY beam test. It can be clearly seen that the current gradually rises over time and this is due to the temperature increase as the dry ice evaporates, as previously described. There is also a large spike in the current at 90 hours which indicates that the sensor broke. Figure~\ref{fi:Current}b shows the current of a IHEP-IMEv2 W4-II 1$\times$3 sensor as a function of time measured during the SPS beam test. The current stays steady over time in comparison to the sensor measured at DESY due to the stable temperature. In both plots, each pause in the current is due to the HV being switched off for reasons such as replacing the dry ice (at DESY only) or removing a broken sensor from the same HV channel. Each large jump in current is caused by the bias being stepped up as part of the testing plan. 

\begin{figure}[t]
\centering
\includegraphics[width=0.8\textwidth]{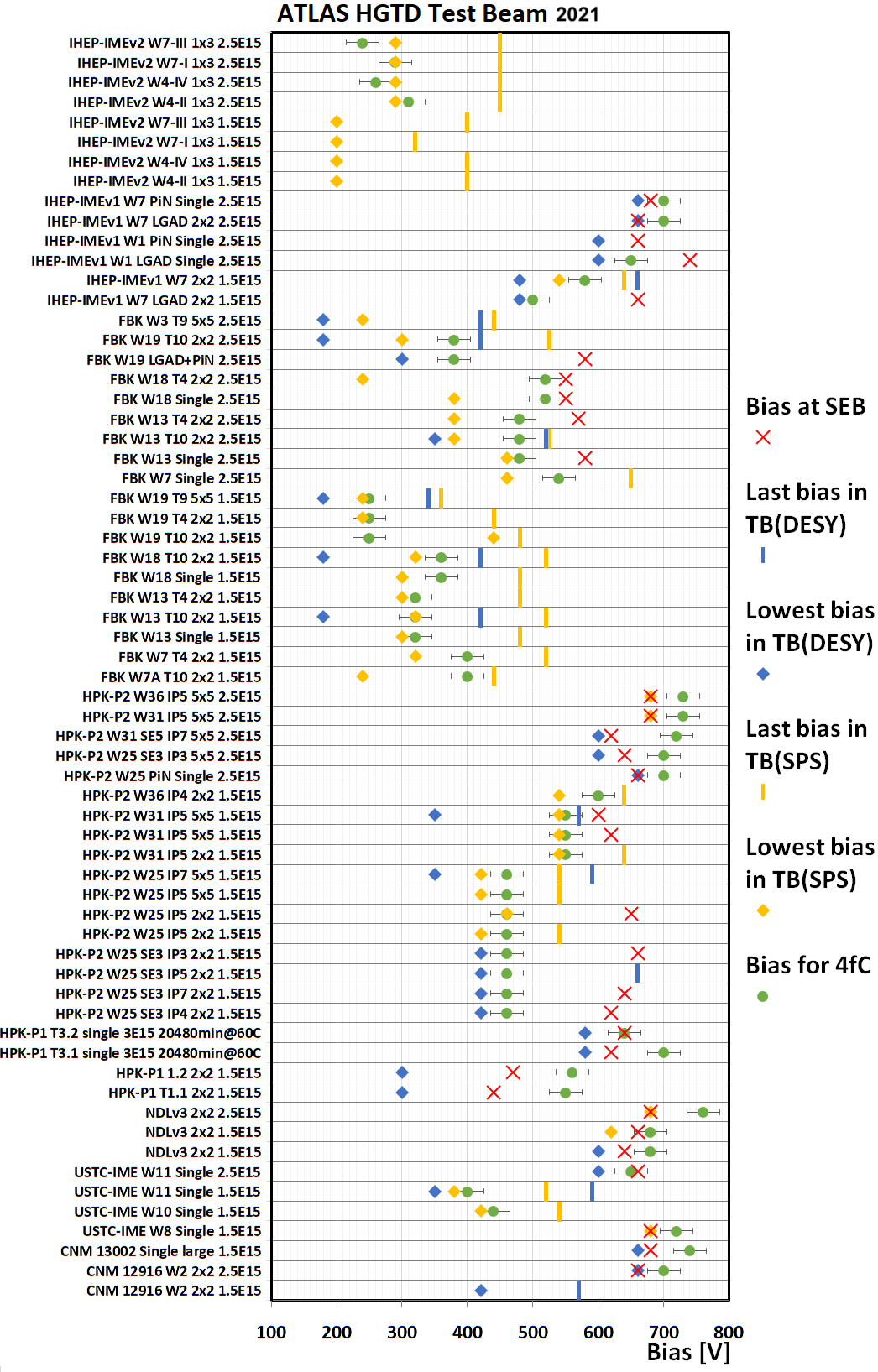}
\caption{Results of both the DESY and SPS beam tests. Green points mark the bias required to reach \SI{4}{\femto\coulomb} and includes the error. The diamond marks are the lowest bias tested while the line marks are the highest bias tested. Yellow marks the measurement range performed at SPS, while blue marks the measurement range performed at DESY. Red crosses mark the bias at which a sensor broke in the beam. Sensors are grouped up by producer.}
\label{fi:Results}
\end{figure}

The results of both beam tests are represented in figure~\ref{fi:Results}. The
graph shows the bias voltage required for the minimum required charge of \SI{4}{\femto\coulomb} (when known
from strontium-90 measurements in the laboratories), the lowest bias tested,
and the highest bias tested. For the sensors that broke it also marks the bias
at which it failed. Based on these results, the sensors that are able to reach the
bias required for \SI{4}{\femto\coulomb} are HPK-P2 W25 and USTC-IME
irradiated to $1.5\times10^{15}$~n$_{eq}$/cm$^{2}$, and IHEP-IMEv2 and FBK-UFSD3.2
irradiated to both $1.5\times10^{15}$~n$_{eq}$/cm$^{2}$ and $2.5\times10^{15}$~n$_{eq}$/cm$^{2}$. The USTC-IME-1.1 wafer 11, IHEP-IMEv2 and FBK sensors are carbon-enriched. None of the NDL nor CNM sensors reached the required bias. 

\begin{figure}[t]
\centering
\begin{tabular}{cc}
\epsfig{file=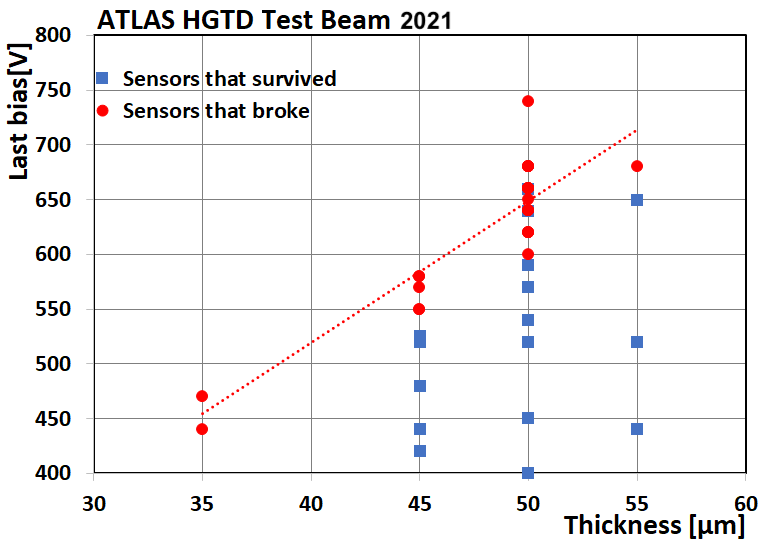,width=0.5\linewidth,clip=} & \epsfig{file=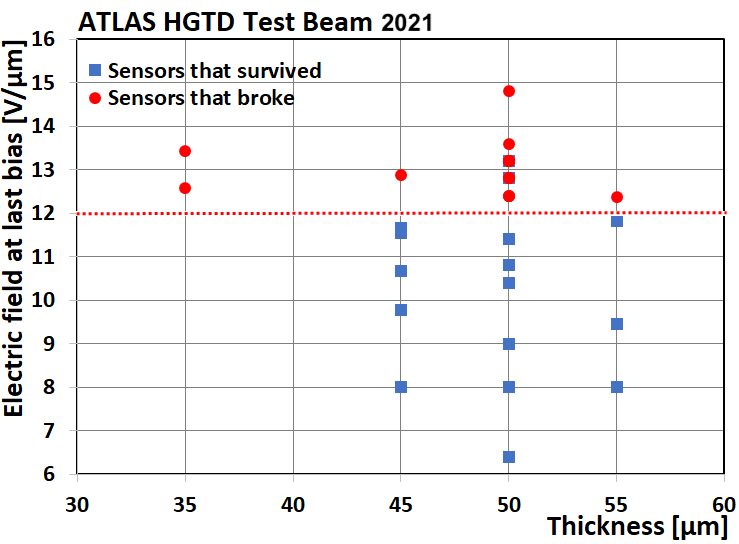,width=0.5\linewidth,clip=} \\
(a) & (b) 
\end{tabular}
\caption{Comparison of (a) the thickness with the last tested bias for all sensors, and (b) the thickness with the electric field in the sensor. In both plots the red circles mark sensors that broke and the red dashed line indicates \SI{12}{\volt\per\micro\metre}.}
\label{fi:BiasVThick}
\end{figure}

In order to determine the conditions under which a sensor breaks in the beam, various active material thicknesses are compared against the last bias and the electric field, both shown in figure~\ref{fi:BiasVThick}.


 As can be seen, sensors with a larger active material thickness were able to
 withstand a higher bias, and when an average field in the sensors is
 calculated, <E>=V$_{bias}$/D, they start to break once they reach
 \SI{12}{\volt\per\micro\metre} regardless of the LGAD design. This explains
 why the sensor from FBK-UFSD3.2 wafer 7 (active thickness \SI{55}{\micro\metre}) irradiated to
 $2.5\times10^{15}$~n$_{eq}$/cm$^{2}$ survived at biases well above
 \SI{600}{\volt} while sensors from other FBK wafers (active thickness
 \SI{45}{\micro\metre}) irradiated to
 $2.5\times10^{15}$~n$_{eq}$/cm$^{2}$ did not. No fatality was
 observed at E<\SI{12}{\volt\per\micro\metre}, although this observation is
 with a limited number of particles crossing the detector. However, to account
 for uncertainty and the fact that breakdown depends exponentially on the
 field, a safe zone of operation of E<\SI{11}{\volt\per\micro\metre} is
 proposed. 

\begin{figure}[t]
\centering
\includegraphics[width=1\textwidth]{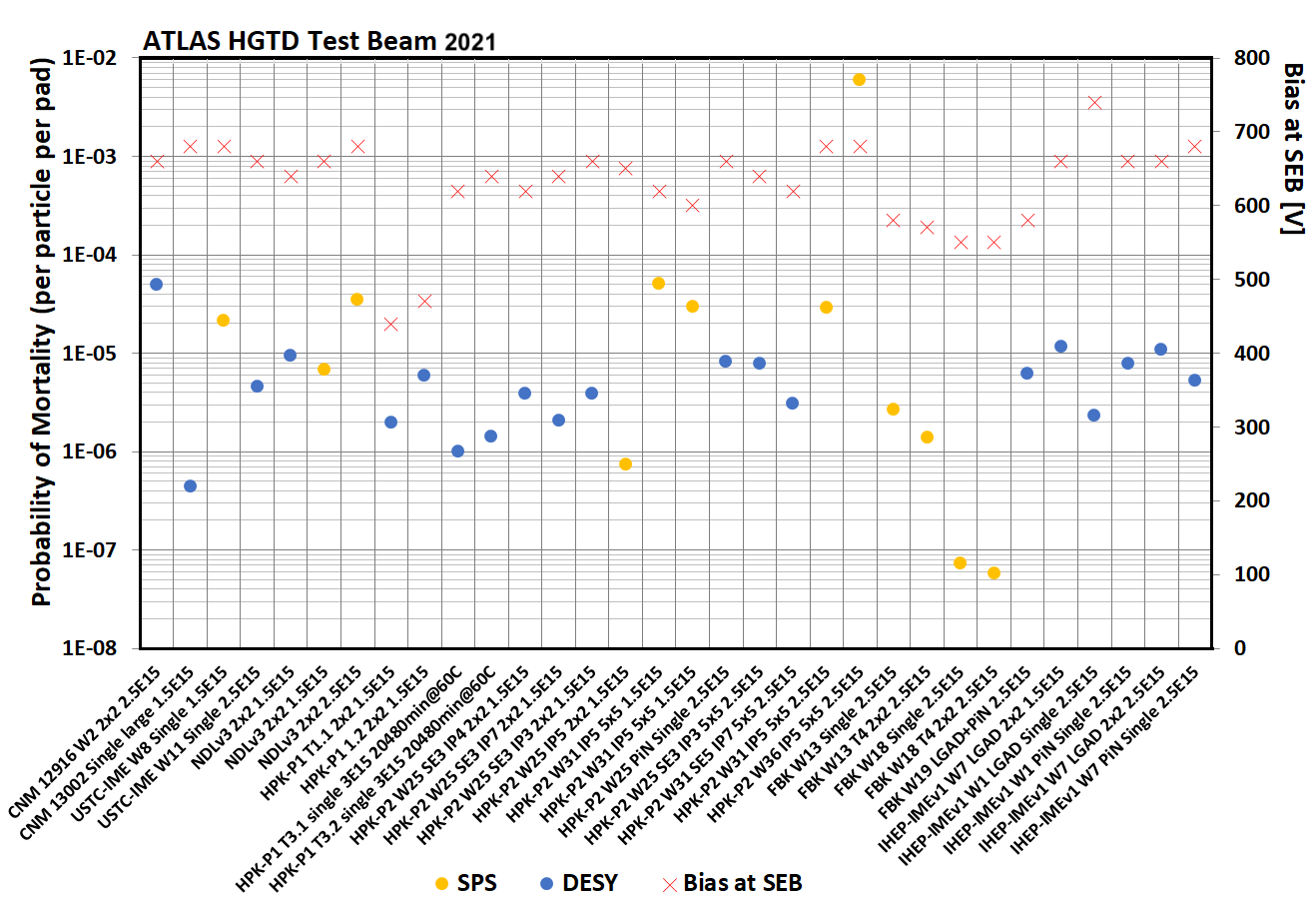}
\caption{Probability of mortality for each particle crossing a single-pad of the sensor, calculated for the sensors that broke at DESY and SPS beams. The bias that the sensor was operated at when mortality occurred is denoted by a red cross.}
\label{fi:prob}
\end{figure}

For the sensors that did not survive, the probability of mortality per
particle crossing the sensor can be calculated. The number of particles
crossing the \SI{1}{\centi\metre\squared} scintillator was recorded, and this
value was scaled down to the number of particles that crossed a single-pad of
each sensor. The probability was then calculated as the inverse of the number
of particles that passed through the sensor pad before it broke.
The precision with which the rate, and therefore the number of particles, is known is better than 50\% due to the control of the beam profile and the efficiency of the particle scintillator counter.
The probabilities are presented in figure~\ref{fi:prob} and are generally in the range of $10^{-5}-10^{-6}$. As would be expected, the sensors that broke at a bias of \SI{680}{\volt} have the highest probability of mortality. However, these sensors were not tested at biases below \SI{680}{\volt}, so it is possible that they would have broken at a lower bias with a lower probability.

\FloatBarrier

\section{Conclusion}
\label{sec:conclusion}
In this study, many sensors from various producers and runs were sent to two test beam facilities to have their survivability tested in the particle beam. Some of the sensors broke at biases lower than those tested in the lab, demonstrating that the particle beam itself causes their mortality. It was determined that it was not the design of the LGAD that changed the chances of mortality, but the strength of the electric field. It is concluded that the safe zone of operation is at electric fields below \SI{11}{\volt\per\micro\metre}.

\section*{Acknowledgements}

The authors gratefully acknowledge CERN and the SPS staff for successfully operating the North Experimental Area and for continuous supports to the users.
The measurements leading to these results have been performed at the Test Beam Facility at DESY Hamburg (Germany), a member of the Helmholtz Association (HGF). 
This work was partially funded by MINECO, Spanish Government, under grant
RTI2018-094906-B-C21. This project has received funding from the European
Union's Horizon 2020 research and innovation programme under the Marie
Sklodowska-Curie grant agreement No. 754510. This work was partially funded
by: the Spanish Government, under grant FPA2015-69260-C3-2-R and SEV-2012-0234
(Severo Ochoa excellence programme); and by the H2020 project AIDA-2020, GA
no. 654168. The authors acknowledge the financial support from the Slovenian
Research Agency (ARRS J1-1699, ARRS P1-0135). The USP group acknowledges
support from FAPESP grant 2020/04867-2 and CAPES.

\pagebreak


\bibliographystyle{unsrt}
\bibliography{paper}

\begin{thebibliography}{10}

\bibitem{ATLAS-TDR-31}
{ATLAS Collaboration}.
\newblock {A High-Granularity Timing Detector for the ATLAS Phase-II Upgrade:
  Technical Design Report}, 2020.

\bibitem{PELLEGRINI201412}
G.~Pellegrini et~al.
\newblock {Technology developments and first measurements of Low Gain Avalanche
  Detectors (LGAD) for high energy physics applications}.
\newblock {\em Nucl. Instrum. Meth. A}, 765:12--16, 2014.

\bibitem{MarTorino}
M.~Carulla et~al.
\newblock First 50\,$\mu$m thick LGAD fabrication at CNM.
\newblock 28th RD50 Workshop, Torino, Italy, 2016.

\bibitem{LGADIrrad}
G.~Kramberger et~al.
\newblock {Radiation effects in Low Gain Avalanche Detectors after hadron
  irradiations}.
\newblock {\em JINST}, 10(07):P07006, 2015.

\bibitem{spsfacility}
{CERN SPS} north area,
  \url{http://sba.web.cern.ch/sba/BeamsAndAreas/H6/H6_presentation.html}.

\bibitem{desyfacility}
R.~Diener et~al.
\newblock {The DESY II Test Beam Facility}.
\newblock {\em Nucl. Instrum. Meth. A}, 922:265--286, 2019.

\bibitem{CNMphoto}
S.~Hidalgo.
\newblock LGAD 4 ATLAS, IMB—CNM activities, work presented during a meeting
  within the ATLAS collaboration, attended by IFAE, SCIPP-UCSC, IJS and
  IMB-CNM, May 2019.

\bibitem{Kramberger}
G.~Kramberger.
\newblock LGADs for timing detectors at HL-LHC.
\newblock Presented at CERN detector seminars, CERN, 2021.

\bibitem{Heller}
R.~Heller.
\newblock Studies of LGAD mortality using the fermilab test beam.
\newblock 38th RD50 Workshop, 2021.

\bibitem{SEE}
Gordana La\v{s}tovi\v{c}ka-Medin, Mateusz Rebarz, Gregor Kramberger,
  Ji\v{r}\'\i{} Kroll, Kamil Kropielnicki, Tom\'a\v{s} La\v{s}tovi\v{c}ka,
  Martin Precek, and Jakob Andreasson.
\newblock {Femtosecond laser studies of the Single Event Effects in Low Gain
  Avalanche Detectors and PINs at ELI Beamlines}.
\newblock {\em Nucl. Instrum. Meth. A}, 1041:167321, 2022.

\bibitem{SEB}
G.~La\v{s}tovi\v{c}ka-Medin, G.~Kramberger, M.~Rebarz, J.~Andreasson,
  K.~Kropielnicki, T.~La\v{s}tovi\v{c}ka, and J.~Kroll.
\newblock {A brief overview of the studies on the irreversible breakdown of
  LGAD testing samples irradiated at the critical LHC-HL fluences}.
\newblock {\em JINST}, 17(07):C07020, 2022.

\bibitem{geant4}
S.~Agostinelli et~al.
\newblock {GEANT4--a simulation toolkit}.
\newblock {\em Nucl. Instrum. Meth. A}, 506:250--303, 2003.

\bibitem{Huhtinen:687409}
Mika Huhtinen.
\newblock {Highly ionising events in silicon detectors}.
\newblock Technical report, CERN, Geneva, 2002.

\bibitem{RD48}
G.~Lindstr\"om et~al.
\newblock Radiation hard silicon detectors--developments by the rd48 (rose)
  collaboration.
\newblock {\em Nuclear Instruments and Methods in Physics Research Section A:
  Accelerators, Spectrometers, Detectors and Associated Equipment},
  466(2):308--326, 2001.
\newblock 4th Int. Symp. on Development and Application of Semiconductor
  Tracking Detectors.

\bibitem{Reactor}
Computational analysis of irradiation facilities at the jsi triga reactor.
\newblock {\em Applied Radiation and Isotopes}, 70(3):483--488, 2012.

\bibitem{LGADCarbon}
M.~Ferrero et~al.
\newblock {Radiation resistant LGAD design}.
\newblock {\em Nucl. Instrum. Meth. A}, 919:16--26, 2019.

\bibitem{HPK3-2018/19}
C.~Agapopoulou et~al.
\newblock {Performance in beam tests of irradiated Low Gain Avalanche Detectors
  for the ATLAS High Granularity Timing Detector}.
\newblock {\em JINST}, 17(09):P09026, 2022.

\end{thebibliography}

\clearpage
\appendix
\part*{Appendices}
\addcontentsline{toc}{part}{Appendices}
\section{List of tested sensors}\label{A:SensorList}
The sensors that were tested in both test beam facilities are listed in this table. The sensor name is made up of the producer, run, wafer, and in certain cases other information such as the type or inter-pad (IP) distance. Where the sensor is a PiN and not an LGAD, this is mentioned in parenthesis next to the sensor name. The size of the sensor refers to the number of pads. In the case of LGAD-PiN this means that there were two pads, with one being an LGAD and the other a PiN. The annealing for the majority of the sensors was done at \SI{60}{\degree C}, unless stated otherwise in parenthesis next to the annealing time. 

\begin{table}[h]
\centering
\caption{List of tested sensors from CNM, USTC-IME, NDL and HPK vendors.}
{\footnotesize
\begin{tabular}{|c|c|c|c|c|c|c|}
\hline
\textbf{Sensor name}           & \begin{tabular}{@{}c@{}}\textbf{Pad} \\ \textbf{structure}\end{tabular} & \begin{tabular}{@{}c@{}}\textbf{Active thickness} \\ \textbf{[$\mu$m]}\end{tabular} & \textbf{$V_{gl}$ [V]} & \begin{tabular}{@{}c@{}}\textbf{Fluence} \\ \textbf{[n$_{eq}$/cm$^{2}$]}\end{tabular} & \begin{tabular}{@{}c@{}}\textbf{Annealing} \\ \textbf{time [min]} \\ \textbf{(@\SI{60}{\degree C})}\end{tabular} & \textbf{Tested at} \\ \hline
CNM   12916 W2            & 2$\times$2     & 50    & 13.2   & $1.5\times10^{15}$         & 80                                        & DESY               \\ 
CNM 12916 W2              & 2$\times$2      & 50   & 6.6   & $2.5\times10^{15}$         & 80                                        & DESY               \\ 
CNM   13002               & single large & 55 & unknown  & $1.5\times10^{15}$         & 80                                        & DESY               \\ \hline
USTC-IME-1.1 W8                   & single  & 50    & 15.6   & $1.5\times10^{15}$        & 80                                        & SPS                \\ 
USTC-IME-1.1 W10                  & single   & 50   & 19.1 & $1.5\times10^{15}$         & 80                                        & SPS                \\ 
USTC-IME-1.1 W11                  & single   & 50   & 29.2  & $1.5\times10^{15}$        & 80                                        & DESY/ SPS           \\ 
USTC-IME-1.1 W11                  & single    & 50  & 23.3  & $2.5\times10^{15}$         & 80                                        & DESY               \\ \hline
NDLv3 B14-D3              & 2$\times$2     & 50  & 10.8    & $1.5\times10^{15}$         & 80                                        & DESY               \\ 
NDLv3 B14-D5              & 2$\times$2     & 50  & 10.8    & $1.5\times10^{15}$        & 80                                        & SPS                \\ 
NDLv3 B14-B5              & 2$\times$2    & 50    & unmeasurable   & $2.5\times10^{15}$         & 80                                        & SPS                \\ \hline
HPK-P1   Type1.1 SE2-IP9  & 2$\times$2     & 35  & unknown    & $1.5\times10^{15}$        & 80                                        & DESY               \\ 
HPK-P1 Type1.2 SE3-IP5    & 2$\times$2     & 35  & unknown    & $1.5\times10^{15}$        & 80                                        & DESY               \\ 
HPK-P1   Type3.1 SE5      & single    & 50 & unknown   & $3\times10^{15}$      & 20480                                     & DESY               \\ 
HPK-P1 Type3.2 SE2        & single   & 50  & unknown   & $3\times10^{15}$         & 20480                                     & DESY               \\ 
HPK-P2 W25 SE3-IP4        & 2$\times$2    & 50   & 29.5    & $1.5\times10^{15}$       & 80                                        & DESY               \\ 
HPK-P2 W25 SE3-IP7        & 2$\times$2      & 50  & 29.5   & $1.5\times10^{15}$      & 80                                        & DESY               \\ 
HPK-P2 W25 SE3-IP5        & 2$\times$2      & 50  & 29.5   & $1.5\times10^{15}$       & 80                                        & DESY               \\ 
HPK-P2 W25 SE3-IP3        & 2$\times$2    & 50   & 29.5    & $1.5\times10^{15}$        & 80                                        & DESY               \\ 
HPK-P2 W25 SE3-IP5        & 2$\times$2    & 50  & 29.5     & $1.5\times10^{15}$    & 80                                        & SPS                \\ 
HPK-P2 W25 SE3-IP5        & 2$\times$2    & 50   & 29.5    & $1.5\times10^{15}$       & 80                                        & SPS                \\ 
HPK-P2 W25   SE3-IP5      & 5$\times$5  & 50    & 29.5   & $1.5\times10^{15}$   & 80                                        & SPS                \\ 
HPK-P2 W25 SE3-IP7        & 5$\times$5    & 50   & 29.5   & $1.5\times10^{15}$      & 80                                        & DESY/ SPS           \\ 
HPK-P2 W31   SE3-IP5      & 2$\times$2   & 50   & 27.4     & $1.5\times10^{15}$     & 80                                        & SPS                \\ 
HPK-P2 W31   SE3-IP5      & 5$\times$5    & 50    & 27.4     & $1.5\times10^{15}$       & 80                                        & SPS                \\ 
HPK-P2 W31 SE3-IP5        & 5$\times$5   & 50     & 27.4     & $1.5\times10^{15}$       & 80                                        & DESY/ SPS           \\ 
HPK-P2 W36 SE3-IP4        & 2$\times$2   & 50    & 24.8     & $1.5\times10^{15}$        & 80                                        & SPS                \\ 
HPK-P2   W25 (PIN)        & single  & 50   & N/A    & $2.5\times10^{15}$       & 80                                        & DESY               \\ 
HPK-P2 W25 SE3-IP3        & 5$\times$5    & 50    & 18.9    & $2.5\times10^{15}$        & 80                                        & DESY               \\ 
HPK-P2   W31 SE5-IP7      & 5$\times$5   & 50  & 18.9       & $2.5\times10^{15}$         & 80                                        & DESY               \\ 
HPK-P2 W31 SE3-IP5        & 5$\times$5   & 50    & 18.9     & $2.5\times10^{15}$        & 80                                        & SPS                \\ 
HPK-P2 W36   SE3-IP5      & 5$\times$5    & 50    & 15.2   & $2.5\times10^{15}$         & 80                                        & SPS                \\ 
\hline 
\end{tabular}
}
\label{tab:sensors1}
\end{table}

\begin{table}[h]
\centering
\caption{List of tested sensors from FBK and IHEP-IME vendors.}
{\footnotesize
\begin{tabular}{|c|c|c|c|c|c|c|}
\hline
\textbf{Sensor name}           & \begin{tabular}{@{}c@{}}\textbf{Pad} \\ \textbf{structure}\end{tabular} & \begin{tabular}{@{}c@{}}\textbf{Thickness} \\ \textbf{[$\mu$m]}\end{tabular} & \textbf{$V_{gl}$ [V]} & \begin{tabular}{@{}c@{}}\textbf{Fluence} \\ \textbf{[n$_{eq}$/cm$^{2}$]}\end{tabular} & \begin{tabular}{@{}c@{}}\textbf{Annealing} \\ \textbf{time [min]} \\ \textbf{(@\SI{60}{\degree C})}\end{tabular} & \textbf{Tested at} \\ \hline
FBK-UFSD3.2   W7A-Type 10 & 2$\times$2   & 55   & 18.0     & $1.5\times10^{15}$      & 80                                        & SPS                \\ 
FBK-UFSD3.2   W7-Type 4   & 2$\times$2      & 55  & 19.6   & $1.5\times10^{15}$        & 80                                        & SPS                \\ 
FBK-UFSD3.2 W13           & LGAD-PiN   & 45 & 34.2 & $1.5\times10^{15}$      & 80                                        & SPS                \\ 
FBK-UFSD3.2 W13-Type 10   & 2$\times$2   & 45  & 34.2     & $1.5\times10^{15}$         & 80                                        & DESY/ SPS           \\ 
FBK-UFSD3.2   W13-Type 4  & 2$\times$2    & 45   & 34.2    & $1.5\times10^{15}$         & 130                                       & SPS                \\ 
FBK-UFSD3.2 W18           & LGAD-PiN   & 45 & 33.1  & $1.5\times10^{15}$        & 80                                        & SPS                \\ 
FBK-UFSD3.2 W18-Type 10   & 2$\times$2      & 45 & 33.1    & $1.5\times10^{15}$         & 80                                        & DESY/ SPS           \\ 
FBK-UFSD3.2 W19-Type 10   & 2$\times$2     & 45   & 37.3   & $1.5\times10^{15}$       & 80                                        & SPS                \\ 
FBK-UFSD3.2   W19-Type 4  & 2$\times$2    & 45   & 37.3    & $1.5\times10^{15}$         & 80                                        & SPS                \\ 
FBK-UFSD3.2 W19-Type 9    & 5$\times$5    & 45   & 37.3    & $1.5\times10^{15}$        & 80                                        & DESY/ SPS           \\ 
FBK-UFSD3.2 W7            & LGAD-PiN    & 55 & 17.6 & $2.5\times10^{15}$       & 80                                        & SPS                \\ 
FBK-UFSD3.2 W13           & LGAD-PiN   & 45 & 29.1   & $2.5\times10^{15}$      & 80                                        & SPS                \\ 
FBK-UFSD3.2 W13-Type 10   & 2$\times$2     & 45  & 29.1    & $2.5\times10^{15}$        & 80                                        & DESY/ SPS           \\ 
FBK-UFSD3.2   W13-Type 4  & 2$\times$2     & 45  & 29.1    & $2.5\times10^{15}$       & 80                                        & SPS                \\ 
FBK-UFSD3.2 W18           & LGAD-PiN   & 45 & 28.5  & $2.5\times10^{15}$       & 80                                        & SPS                \\ 
FBK-UFSD3.2   W18-Type 4  & 2$\times$2      & 45  & 28.5   & $2.5\times10^{15}$     & 80                                        & SPS                \\ 
FBK-UFSD3.2   W19         & LGAD-PiN & 45  & 32.4   & $2.5\times10^{15}$        & 80                                        & DESY               \\ 
FBK-UFSD3.2   W19-Type 10 & 2$\times$2     & 45  & 32.4  & $2.5\times10^{15}$        & 80                                        & DESY/ SPS           \\ 
FBK-UFSD3.2 W3-Type 9     & 5$\times$5    & 45  &  unknown     & $2.5\times10^{15}$       & 80                                        & DESY/ SPS           \\ \hline
IHEP-IMEv1 W7                  & 2$\times$2   & 50    & 14.3       & $1.5\times10^{15}$        & 80                                        & DESY               \\ 
IHEP-IMEv1 W7                  & 2$\times$2   & 50    & 14.3    & $1.5\times10^{15}$         & 80                                        & DESY/ SPS           \\ 
IHEP-IMEv1 W1                  & single   & 50  & 10.8   & $2.5\times10^{15}$        & 80                                        & DESY               \\ 
IHEP-IMEv1 W1 (PIN)              & single    & 50  & N/A   & $2.5\times10^{15}$       & 80                                        & DESY               \\ 
IHEP-IMEv1 W7                  & 2$\times$2   & 50  & 11.1   & $2.5\times10^{15}$        & 80                                        & DESY               \\ 
IHEP-IMEv1 W7 (PIN)            & single   & 50   &N/A   & $2.5\times10^{15}$         & 80                                        & DESY               \\ 
IHEP-IMEv2 W4-II               & 1$\times$3      & 50  & 18.3    & $1.5\times10^{15}$        & 80                                        & SPS                \\ 
IHEP-IMEv2 W4-IV               & 1$\times$3   & 50   & 19.7     & $1.5\times10^{15}$        & 80                                        & SPS                \\ 
IHEP-IMEv2 W7-I                & 1$\times$3    & 50   & 18.7    & $1.5\times10^{15}$       & 80                                        & SPS                \\ 
IHEP-IMEv2 W7-III              & 1$\times$3   & 50  & 20.3     & $1.5\times10^{15}$        & 80                                        & SPS                \\ 
IHEP-IMEv2 W4-II               & 1$\times$3    & 50   & 16.7    & $2.5\times10^{15}$       & 80 (@90$^\circ$C)                         & SPS                \\ 
IHEP-IMEv2 W4-IV               & 1$\times$3   & 50   & 17.2     & $2.5\times10^{15}$        & 80 (@90$^\circ$C)                         & SPS                \\ 
IHEP-IMEv2 W7-I                & 1$\times$3    & 50   & 16.3    & $2.5\times10^{15}$        & 80 (@90$^\circ$C)                         & SPS                \\ 
IHEP-IMEv2 W7-III              & 1$\times$3   & 50    & 18.7    & $2.5\times10^{15}$      & 80 (@90$^\circ$C)                         & SPS                \\
\hline 
\end{tabular}
}
\label{tab:sensors2}
\end{table}

\end{document}